\documentclass[10pt,twocolumn,letterpaper]{article}

\usepackage[pagenumbers]{cvpr} 

\usepackage[dvipsnames]{xcolor}

\definecolor{cvprblue}{rgb}{0.21,0.49,0.74}
\usepackage{graphicx}
\usepackage{nicematrix,enumitem}
\usepackage{booktabs}
\usepackage{multirow}
\usepackage[pagebackref,breaklinks,colorlinks,citecolor=cvprblue]{hyperref}

\def\paperID{17} 
\def\confName{CVPR}
\def\confYear{2024}

\title{Optimizing Synthetic Correlated Diffusion Imaging for Breast Cancer Tumour Delineation}

\author{Chi-en Amy Tai \\
University of Waterloo\\
Waterloo, ON\\
{\tt\small amy.tai@uwaterloo.ca}
\and Alexander Wong \\
University of Waterloo\\
Waterloo, ON\\
{\tt\small alexander.wong@uwaterloo.ca}
}

\begin{document}
\maketitle
\begin{abstract}
Breast cancer is a significant cause of death from cancer in women globally, highlighting the need for improved diagnostic imaging to enhance patient outcomes. Accurate tumour identification is essential for diagnosis, treatment, and monitoring, emphasizing the importance of advanced imaging technologies that provide detailed views of tumour characteristics and disease. Synthetic correlated diffusion imaging (CDI\textsuperscript{s}) is a recent method that has shown promise for prostate cancer delineation compared to current MRI images. In this paper, we explore tuning the coefficients in the computation of CDI\textsuperscript{s} for breast cancer tumour delineation by maximizing the area under the receiver operating characteristic curve (AUC) using a Nelder-Mead simplex optimization strategy. We show that the best AUC is achieved by the CDI\textsuperscript{s} - Optimized modality, outperforming the best gold-standard modality by 0.0044. Notably, the optimized CDI\textsuperscript{s} modality also achieves AUC values over 0.02 higher than the Unoptimized CDI\textsuperscript{s} value, demonstrating the importance of optimizing the CDI\textsuperscript{s} exponents for the specific cancer application. 
\end{abstract}    
\section{Introduction}
\label{sec:intro}
Breast cancer remains a significant cause of cancer-related mortality for women worldwide~\cite{cancer-cases-stats}, necessitating advancements in diagnostic imaging techniques to improve patient outcomes. The precise delineation of breast cancer tumours is crucial for accurate diagnosis, treatment planning, and monitoring~\cite{akakuru2022chemotherapeutic}, underscoring the need for innovative imaging methodologies that can offer detailed insights into tumour morphology and pathology~\cite{mahmoud2023delineation}.

Synthetic correlated diffusion imaging (CDI\textsuperscript{s}) is a recent method that has shown promise for prostate cancer delineation~\cite{wong2022synthetic} compared to current MRI images.  CDI\textsuperscript{s} is a technique that builds on top of correlated diffusion imaging (CDI). CDI analyzes the direction of diffusion in the cancerous tissue whereas CDI\textsuperscript{s} introduces synthetic signals by extrapolating MRI data to introduce more data points~\cite{wong2022synthetic}. The process begins with multiple native DWI signals obtained for different b-value which are then passed into a signal synthesizer which produces synthetic signals. The native signals are then mixed with the synthetic signals to obtain the CDI\textsuperscript{s} signal~\cite{wong2022synthetic}. Though CDI\textsuperscript{s} served as a strong indicator for prostate presence in tissue~\cite{wong2022synthetic}, there exists a few challenges for implementing CDI\textsuperscript{s} for breast cancer.

As defined in~\cite{wong2022synthetic}, there are two key components for computing CDI\textsuperscript{s}: (1) the calibrated signal mixing function and (2) synthetic signal acquisitions which are mixed with native signal acquisitions. The first component uses $\rho$, which are coefficients that control the contribution of different gradient pulse strengths and timings to produce the CDI\textsuperscript{s} signal. The second component relies on defining ${\hat{S}}$, the specific synthetic signals to acquire. However, the challenge of what values to use for $\rho$ and ${\hat{S}}$ are non-trivial as these values largely impact the quality of the CDI\textsuperscript{s} signal. Selecting optimal parameters by hand is not only labor-intensive but also time-intensive, making it advantageous to identify a strategy for optimizing these parameters for the specific task. Notably, it is also important that $\rho$ does not have values that are too large as the mixing function used in computing CDI\textsuperscript{s} combines signals multiplicatively and overflow errors would occur if the $\rho$ values are too high. 

In this paper, we tune the coefficients in the computation of CDI\textsuperscript{s} for breast cancer tumour delineation by maximizing the area under the receiver operating characteristic curve (AUC) using a Nelder-Mead simplex optimization strategy. 

\section{Methodology}
\label{sec:methodology}

\begin{figure}
  \centering
   \includegraphics[width=\linewidth]{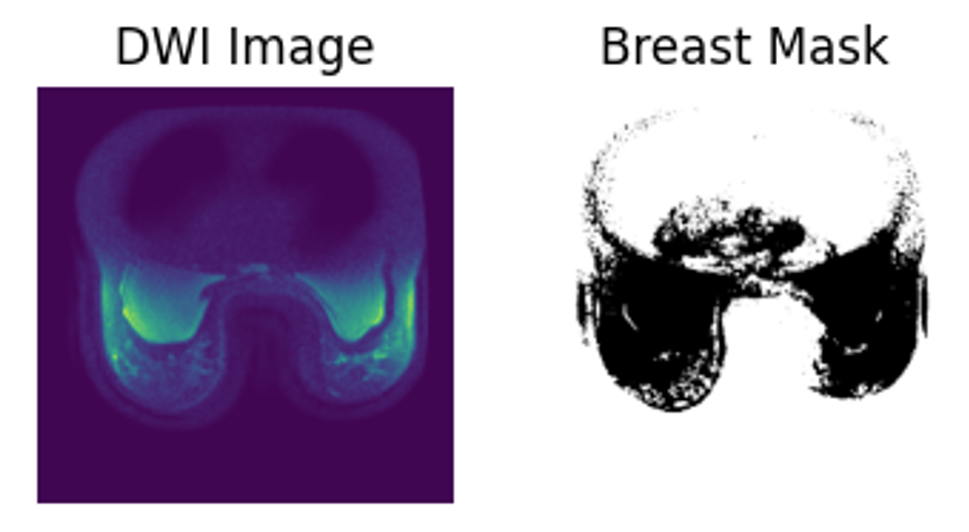}
   \caption{Sample breast mask from the processed DWI image.}
   \label{fig:sample-dwi-breast-mask}
\end{figure}

We use the ACRIN 6698/I-SPY2 study, which contains DWI acquisitions, ADC maps, and DWI whole-tumor segmentations across 10 different institutions for 355 patients at four different timepoints in their treatment~\cite{acrin6698-data-1,acrin6698-data-2,acrin6698-data-3,acrin6698-data-4}. We standardize the MRI images and reduce each MRI volume to 25 slices, the minimum number of slices across all patients, and resize each image in the volume to 224 x 224. DWI volumes also had an extra dimension for the specific b-value. Notably, for comparison, only the DWI images corresponding to a b-value of 800 were considered as a previous study showed that only using this b-value gained higher performance than using other b-values and the b-value of 800 was also more accurate than trying to feed in the data for all the b-values for a similar prediction task~\cite{tai2022cancer}. All b-values are used for the optimization of CDI\textsuperscript{s}. One patient had to be removed as they only had three b-values instead of four so 354 patients were used in this study. As the ACRIN 6698/I-SPY2 study does not contain breast masks, 
To compute the breast mask, thresholding on the processed DWI images were leveraged along with manual inspection of the resulting breast segmentation mask for quality. An example of the generated mask is provided in Figure~\ref{fig:sample-dwi-breast-mask}.

For completeness, a computed form of ADC, ADCc is also calculated based on the specific DWI image. The difference between ADC and ADCc is that ADC is provided in the dataset whereas ADCc is computed manually using a different technique. From the dataset, ADC is calculated as a linear fit to $log(S(b)) = log(S(0)) - ADC * b$ with thresholding as defined in~\cite{acrin6698-data-3}. On the other hand, ADCc uses a standard linear regression and linear least-squares formulation with a $R^2$ threshold of 0.8 to compute the ADC map.

We leverage the optimization setup from~\cite{wong2022synthetic} with an initial $\rho$ of [1.6160, 1.5209, 1.2006, 0.8362, 1.1630, 0.8666, 1.1424, -0.4635] (showing rounded values for brevity), and ${\hat{S}}$ values of [50, 1000, 2000, 3000, 4000, 5000, 6000, 7000]. We also added bounds of [-10, 10] for $\rho$ in the optimization to avoid overflow errors when computing CDI\textsuperscript{s}. Similar to~\cite{wong2022synthetic}, the Nelder-Mead simplex optimization strategy was used to maximize the AUC for ability to delineate between healthy and cancerous breast tissue~\cite{nelder1965simplex}. Though the Nelder-Mead simplex method is not guaranteed to find the global minimum, it was chosen for its computational efficiency and highly opportunistic behaviour~\cite{nelder1965simplex}. The results from these two setups were compared to the unoptimized form that uses the base $\rho = [1, 1, 1, 1, 1, 1]$ values and ${\hat{S}} = [0, 1000, 2000, 3000, 4000, 5000]$, provided in~\cite{tai2023multi}.

\section{Results}
\label{sec:results}

\begin{figure}
  \centering
   \includegraphics[width=\linewidth]{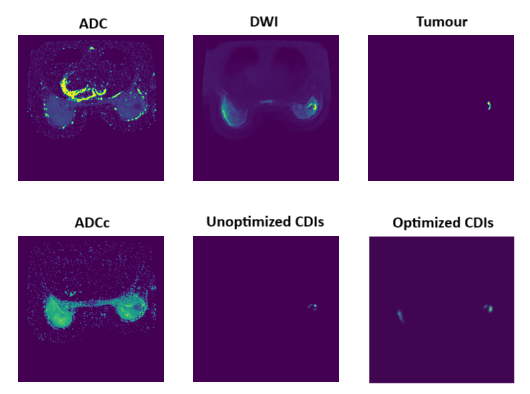}
   \caption{Visual comparison of the tumour mask, ADC, DWI, ADCc, Unoptimized CDI\textsuperscript{s}, and Optimized CDI\textsuperscript{s}.}
   \label{fig:sample-original-modality}
\end{figure}

Figure~\ref{fig:sample-original-modality} shows the visual comparison of the tumour mask and the MRI image modalities (ADC, DWI, ADCc, Unoptimized CDI\textsuperscript{s}, and Optimized CDI\textsuperscript{s}). As seen in Figure~\ref{fig:sample-original-modality}, Optimized CDI\textsuperscript{s} is able to best capture the tumour region with the least amount of noise compared to the other modalities. 

Table~\ref{tab:modalities-auc} shows the AUC values for the various modalities to separate healthy and tumour tissue. The best AUC value on the processed images is achieved by the Optimized CDI\textsuperscript{s} modality, outperforming the best gold-standard modality by 0.0044. Notably, the optimized CDI\textsuperscript{s} modality also achieves AUC values over 0.02 higher than the Unoptimized CDI\textsuperscript{s} value, demonstrating the importance of optimizing the CDI\textsuperscript{s} exponents for the specific cancer application. 

Although these results are promising, the optimization of CDI\textsuperscript{s} was conducted using basic threshold-derived breast masks from the DWI images and were not verified by experienced radiologists. Moreover, the performance improvement of CDI\textsuperscript{s} over the best gold-standard imaging modality is marginal and could differ for another dataset. Though the Nelder-Mead optimization strategy is widely used, there is still the possibility that the chosen optimization coefficients were not globally optimal and there exists better coefficients which could be used. Lastly, since tumour masks were not provided for T2w images, the AUC performance could not be computed for the T2w modality, another gold-standard imaging modality, which may be able to better separate healthy and tumour tissue for breast.

\begin{table}
    \caption{AUC values for the various modalities to separate healthy and tumour tissue with highest AUC value indicated by bold.}
    \centering
    \begin{NiceTabular}{l c}
    \toprule
    \RowStyle{\bfseries}
    \textbf{Modality} & \textbf{AUC} \\ \midrule
    ADC & 0.8323 \\
    DWI & 0.9426 \\
    ADCc & 0.8847 \\
    Unoptimized CDI\textsuperscript{s} & 0.9224 \\
    Optimized CDI\textsuperscript{s} & 0.9470 \\
    \end{NiceTabular}
    \label{tab:modalities-auc}
\end{table}

{
    \small
    \bibliographystyle{ieeenat_fullname}
    \bibliography{main}
}

\end{document}